# Peer-to-Peer and Mass Communication Effect on Revolution Dynamics

Kindler A[1,¶], Solomon S[1], Stauffer D[2]

[1] Racah Institute of Physics, Hebrew University of Jerusalem, Jerusalem, Israel

[2] Institute for Theoretical Physics, Cologne University, D-50923 Köln, Euroland

[¶] Corresponding author: alexanderkindler@gmail.com

*Abstract*

*Revolution dynamics is studied through a minimal Ising model with three main influences (fields): personal conservatism (power-law distributed), inter-personal and group pressure, and a global field incorporating peer-to-peer and mass communications, which is generated bottom-up from the revolutionary faction. A rich phase diagram appears separating possible terminal stages of the revolution, characterizing failure phases by the features of the individuals who had joined the revolution. An exhaustive solution of the model is produced, allowing predictions to be made on the revolution's outcome.*

## 1.1 Introduction

It was always the nature of revolutions that they started from rather localized events, but developed due to systemic conditions.

Examples escort the human history from the time of the discovery of fire and agriculture, through the transformation of the Goths from a band of refugees into the dominant force in Europe, and to the current social and economical crises.

The scientific quantitative study of such phenomena has been envisaged in the last decades but it is significant that some of the trials were formulated at the fringe of the scientific community (e.g. the fictional science Psychohistory, introduced by Isaac Asimov in his Foundation universe or the reflexivity idea formulated by Soros in the 60's but published only recently [1]).

The recent capabilities, afforded by the internet in peer-to-peer and in enhanced mass communication, have further enhanced the possibility of individuals to influence not only their limited number of acquaintances, but also a significantly larger amount of individuals. This has been shown to completely change not only the conditions but also the character of the transition that the system, i.e. network of people, undergoes [2].

In the present paper, we consider the minimal dynamical model that describes the revolution dynamics in the presence of peer-to-peer and mass communication. In doing so, we include both the new elements introduced by the internet communication (communication unlimited by physical distance) as well as elements that characterize the human society since pre-history (mutual influence between close individuals, the reactionary influence of the establishment).

We include three effects which influence revolutionary diffusion in our model: (1) personal affiliation towards change, (2) group pressure of closest acquaintances, and (3) global influence, such as media, peer-to-peer and mass communication (especially Internet and cellular oriented peer-to-peer devices, such as: phone calls, SMS, emails, twitter, Internet sites and blogs, etc). The global influence (equation (1)) is created bottom-up from the fraction of change adopters, and then acts top-down on the entire population.

The ideas above define the core of our model, and understanding their interdependence in relation to the outcome of the revolution is our primary goal. Previous works in the field (for example[3], [4]) include only some of the effects described above, but not a full combination of them.

Our results imply that the degree of the revolutionary success may be predicted. In other words, the influence of societal characteristics in respect to revolutionary parameters is clarified.

### 1.2 Method

It has been previously [5] postulated, that social mass-behavior might be modeled through the Ising model, whereas the value of each spin $i$ represents the state adopted by that individual in respect to his revolutionary view. Given that, we examine when total revolution takes place, as inspired percolation theory [6], [7].

In the present paper, we have initially used an a-thermal 2D square lattice of $L \times L$ spins. In a more general case of network structure, $L$ may take different meanings, and could even lose its relevance altogether. The total number of individuals is defined as $N_{tot}$, and in our specific implementation $N_{tot} = L^2$.

A square lattice compels high probabilities that closest acquaintances of any individual may have another common closest acquaintance between them. This signifies a certain structure of the society.

We have also used a random graph model with similar characteristics to the square-lattice, where each individual (node) has exactly four neighbors, but their geographical representation is random, so that the probability that the closest acquaintances of any individual may have a common closest acquaintance between them is very small. In addition, the graph is fully connected (no unlinked components), to focus on the major part of the society engaged in the revolution. At any rate, following the small-world experiments [8], [9], [10] the graph should not allow unlinked components, and thus should be fully connected. The graph generation process is similar to that proposed in [11].

Let us express the elements of the model quantitatively.

1. We express the overt position of each individual $i$ by a variable $S_i$, that is +1 in the case that it conforms to the "old" order and -1 if it adopts the new "revolutionary" stand.

2. We take into account that the "loyalty" of each individual $i$ to the establishment is a heterogeneous variable, which we label $h_i>0$.
    a. For somebody who is completely immune to the influence of the establishment $h_i=0$, while for somebody who is very attached/ obedient/ aligned with the old rule, $h_i$ is large.
    b. In agreement with many empirical facts from similar systems [12], [13], we assume that the distribution of values $h_i$ is given by a power law: the fraction of individuals $i$ whose initial attachment to the old order is less than $h$ is given by the cumulative probability distribution $P(h_i < h) \propto h^c$ for $h \in [0, d]$. $P(h_i < h) = 1$ if $h>d$, while $d$ and $c$ are parameters. The normalization explicitly demands $P(h_i < h) = (h/d)^c$.

3. We assume that only several ($N_s > 0$) individuals across the network are initially revolutionary predisposed, and thus for them $S=-1$. These individuals are randomly distributed (uniformly) and thus are not necessarily spatially related. All other individuals are initially aligned with the old rule, and thus for them $S=+1$.

4. All other individuals may change their stand if the global and local pressures as detailed below upset their initial allegiance to the old order.

5. The global influence depends on the total current number $N$ of "revolutionaries" [14]. We will employ now and further in the research the notation of $R=N/N_{tot}$ and $R_{init}=N_s/N_{tot}$ for convenience.

$$(1) \quad H(R) = aR^b \quad \text{where } a \text{ and } b \text{ are parameters.}$$

6. The local influence is assumed to be proportional the sum of the positions of the "acquaintances" of $i$ [defined by the links $(i, j)$ of the social network of $i$]: $H_{local}(i) = J \cdot \sum_j S_j$, where $J$ is a parameter and the acquaintances $j$ of $i$ are defined through assuming a particular social network between the individuals in the system. For example, in the case of a 2D square lattice, each individual has four acquaintances: North, East, West and South.

7. The rules 4, 5, 6 have the potential of insuring the propagation of the "revolution" across the network's links. More precisely, if the condition

$$(2) \quad h_i + H_{local}(i) < H$$

is fulfilled, then $i$ will join the revolution. Our 'minimal' model does not allow individuals to change back their tendencies, and to leave the revolution. This requirement maintains the initial $N_s$ revolutionaries, and does not allow them to be swallowed-up by the conformists.

In order to avoid scaling problems, we have $J \leq 1$, $0 < b < 1$, $c \geq 1$ and without loss of generality we set $d=1$. Besides that, we will work with relatively low values of $R_{init}=N_s/N_{tot}$ (up to 0.1) since if they are large, the meaning of a 'revolution' becomes fickle.

The appendix proffers a brief view into spontaneous revolutions, i.e. when temperature is finite. However, the main focus of this work is the a-thermal model as described above.

## 1.3 Revolution Dynamics

To begin with an even more simplified model, we take $J<<1$, meaning that first-hand relations are a negligible factor in determining a person's political behavior. A more moderate approach would simply assume that the society reacts strongly to media in a certain given issue (politics, purchase of a new gadget, *American Idol* etc).

It is important to note that when $J \ll 1$ the dimensionality and connectivity of the network becomes unimportant, since $H$ is a global field independent of the network's structure, and the number and orientation of close neighbors becomes irrelevant to system dynamics.

With this in mind, we may reach two conditions for revolutionary propagation for $J \ll 1$, as elaborated in [15]. The first is a necessary condition for total revolution, which follows from requiring that at large fractions of revolutionaries, i.e. when $R = N/N_{tot} \approx 1$, the revolution must still propagate:

$$(3a) \quad a \geq 1$$

The second condition for revolutionary success is derived from the requirement of having the initial conditions start a dynamics where $R$ grows until reaching unity:

$$(3b) \quad a = R_{init}^{-b+1/c}$$

Of course, the actual condition on $a$ could be simplified as that satisfying both equations (3a) and (3b), thus concluding with $a = \max\left(1, R_{init}^{-b+1/c}\right)$. Figure 1 agrees with equations (3a) and (3b) which are mathematically examined in [15]:

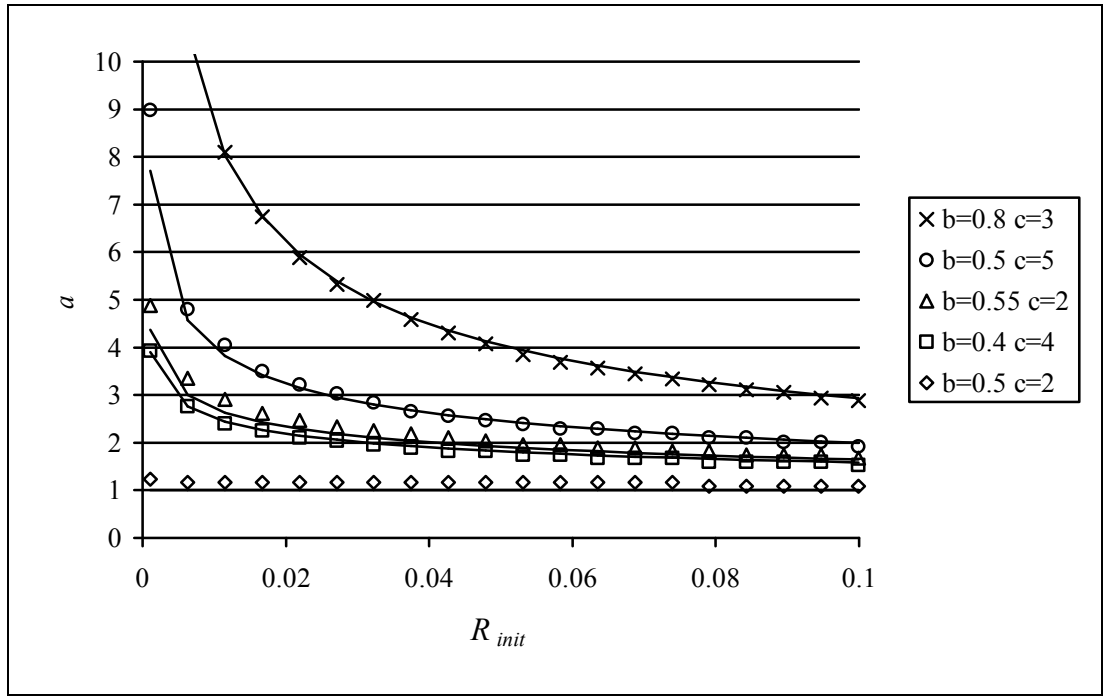

Figure 1: The minimal values of $a$ for which the network undergoes total revolution. Listed are networks with different parameters ($b, c$). The symbols (square, circle, cross, triangle and rhomb) represent simulated values, while the solid lines going through them represent our theoretical prediction according to condition (3a) and (3b).

It is also of interest to investigate how the network evolves when condition (3b) holds, but condition (3a) does not. For example, we will take large values of $R_{init}= N_s/N_{tot}$, but only a slightly smaller value of $a$ than the required $d$. Figure 2 shows the fraction of revolutionaries at the final stage, $R_\infty = \dfrac{N_{final}}{N_{tot}}$, as a function of the fraction of initial revolutionaries, $R_{init} = \dfrac{N_s}{N_{tot}}$. The different curves correspond to different values of $a$ for three different networks, where $a=1$ or $a=0.96$. The networks differ in parameters such as $b$ and $c$. The rather significant differences between the results for low values of $R_{init}$ are due to condition (3b):

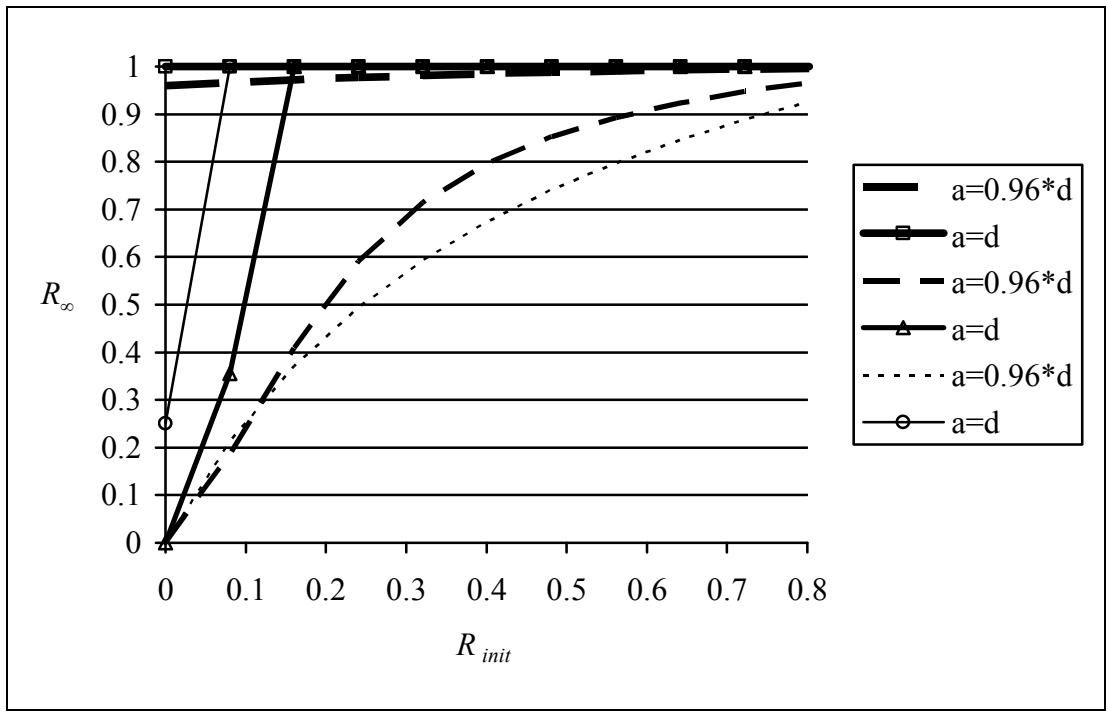

Figure 2: Simulated results of three different networks, presenting the percentage of revolutionaries after the system has evolved to a steady state versus the initial percentage of revolutionaries. All graphs were made for networks with $N_{tot}=90000$. Heavy lines: $c=2$, $b=0.2$. Medium lines: $c=3$, $b=0.3$. Light lines: $c=4$, $b=0.2$. In all three cases: Solid line: $a=1$ [from condition (3a)] to achieve total revolution; dashed line: $a=0.96$ for partial revolution.

We clearly see that even for $a$ very close to 1, as required by condition (3a), and for extremely high values of $R_{init}$, we do not get a total revolution

Now, let us consider a society where individuals are also affected by the opinions of close friends and acquaintances, and not only by the media-communicated public tendency.

The dynamics of the network will be determined as before by equation (2) $h_i + H_{local}(i) < H_0$, though now $H_{local}(i)$ may not be neglected.

For $J>0$, $H_{local}(i)$ can take both negative and positive values as a function of the number of revolutionary neighbors of $i$, and thus $J>0$ may either boost or impede the revolution's spread. Given that, we address the issue through examining the amount of revolutionary neighbors that an old-rule supporter could have.

Moreover, the geographical representation of closest neighbors (or acquaintances) an individual has depends on the type of network we utilize. As formerly stated, we have employed two types of networks for studying our minimal model: a 2D square lattice, and a random graph where each node is of degree 4.

We introduce the fractions $\{Z_k\}$ denoting the number (divided by $N_{tot}$) of old-rule supporters who have $k$ neighboring revolutionaries. In our minimal model $k$ ranges among 0,1,2,3,4.

A straightforward mathematical exercise (elaborated in [15]) produces the following result:

$$(4)\; Z_k(R_{init}) = \binom{4}{k} \cdot R_{init}^k \cdot (1 - R_{init})^{5-k}.$$

For significant values of $J$, the probability of an individual to join the revolution would grow with $k$, where $Z_k$ is the group to which this individual belongs.

In other words, a network should begin its revolution dynamics as determined by equation (3). Then, individuals should join the revolution first from the $Z_4$ and $Z_3$ groups, for whom $J$ enhances the chances of revolting, then from $Z_2$, which is indifferent to $J$, and only later from $Z_1$ and $Z_0$, for whom $J$ hinders the revolution. Of course, this argument is inaccurate if the global field $H$ is much more significant than the local field $H_{local}(i)$ which ranges up to $8J$ (from $-4J$ for $Z_4$ to $4J$ for $Z_0$).

Figure 3 offers a glimpse into the dynamics of the revolution across a network, with respect to the evolvement of $Z_k$ and the growing (time dependent) revolutionary fraction $R_t$. In these graphs, we work with very high $R_{init}$ only for demonstration purposes.

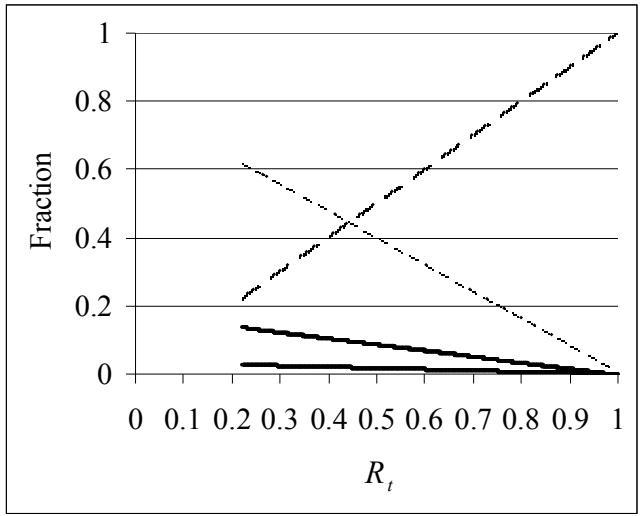

4 (a)

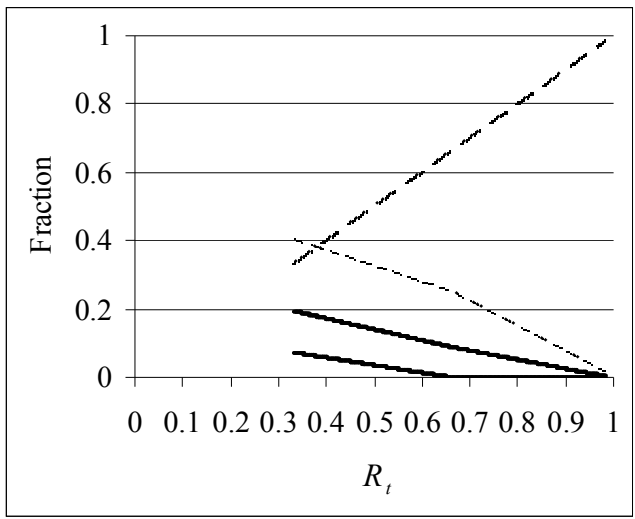

4 (b)

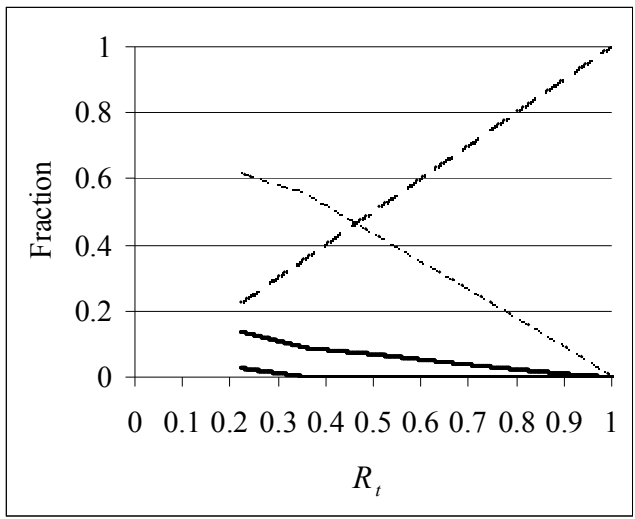

4 (c)

<u>Figure 3:</u> Simulated dynamical behavior of $\{Z_k\}$ in a square lattice of $N_{tot}$=90,000 sites with $b$=0.2, $c$=2, and $J$=1. Heavy-dashed line represents $R_t$ itself for convenience (the growing revolutionary fraction); heavy-solid line represents the $Z_3+Z_4$ fraction, for whom $H_{local}(i)$ enhances the revolution; medium-solid line represents $Z_2$ for whom $H_{local}(i)$=0 is neutral, and the medium-dashed line represents $Z_0+Z_1$ for whom $H_{local}(i)$ inhibits the revolution. $R_{init}$ is 0.22 for parts (a) and (c), and is 0.33 for part (b); $a$=2 for part (a) and $a$=1 for parts (b) and (c).

The behavior in the three graphs is different. In Figure 4a, where $a$ is twice as large as $J$, we witness a steady, uniform decline of all fractions $Z_k$ towards 0 as $R_t$ grows towards unity (total revolution). In this case, $J$ is not sufficiently large to have a significant impact on the dynamics.

In Figure 4b we have $a=J=1$, and we see that the revolution-enhancing fractions ($Z_3+Z_4$) are exhausted before the others, while the revolution-impeding fractions ($Z_0+Z_1$) are more slowly converted into revolutionaries (or other $Z_k$, as is often the case). This means that not all individuals having one or less revolutionary neighbors are able to be among the first to join this revolution (depending of course on individual $h_i$), but rather a stronger global influence must be achieved in order to sweep them away. Those having two or more revolutionary neighbors join the revolution faster.

Figure 4c is a similar case, where at first only those from the ($Z_3+Z_4$) and $Z_2$ are converted, and only later the rest of the population. We can verify this by the stronger slopes in their decline at the beginning, as compared with ($Z_0+Z_1$).

Figure 4 has shown that different possible dynamics may lead to the same end of the revolution, in the discussed cases: total revolution. We will now turn to investigate the different results of a revolutionary spread.

### 1.4 Terminal Phases of a Revolution

A rather interesting inquiry is whether we may characterize a systemic terminal stage of the revolutionary progress better than 'success' or 'failure'. Hence, it is promising to examine the 'failure' results by the features of the individuals who have or have not joined in, and as discussed above, the fractions $Z_k$ would be instrumental in this analysis.

In other words, we examine the distribution of each $Z_k$ at the infinite time limit of the evolution. If all $Z_m, Z_{m+1}, ..., Z_4$ are empty, then $m$ defines the phase.

Let us denote the possible system phases:

A. The network maintains the initial $N_s$ revolutionaries, and no new revolutionaries join.

B. All individuals who at any time are 'surrounded' by 4 revolutionary neighbors join the revolution, but $Z_3$, $Z_2$, $Z_1$, and $Z_0$ remain non-empty when steady-state is reached.

C. Same as B, but now 3 revolutionary neighbors at any time suffice for any individual to join the revolution.

D. Same as B, but now 2 revolutionary neighbors at any time suffice for any individual to join the revolution

E. Same as B, but now a single revolutionary neighbor at any time suffices for any individual to join the revolution.

F. Total revolution – all individuals join the revolution regardless of the number of revolutionary neighbors they may have had.

G. In this phase the final number of revolutionaries is larger than $N_s$ but for all $k=0,1,2,3,4$ the number of individuals belonging to $Z_k$ is larger than zero.

The above phases are an exhaustive description of the system. We should add, that phase G is rather controlled by the distribution of $\{h_i\}$ than by the count of neighboring revolutionaries, and for enough realizations and a diverse distribution of $\{h_i\}$ may encompass the A phase. This is due to the growing probability of obtaining a very low (negligible) $h_i$ and randomly placing it with enough revolutionary neighbors, so that $aR^b > H_{local}(i)$.

Figure 4 shows how these phases depend on different parameters, for a square lattice and a random graph with each node of degree 4. The phase diagrams are presented in a $a$-$J$ plot, having both the global and the local fields change, for invariable 'intrinsic' system parameters: $N_{tot}=90000$ and $c=5000$ (i.e. $h_i \approx 1$ for all $i$) or $c=2$. The diagrams corresponding to $c=5000$ are presented in the left column, and those for whom $c=2$ are in the right column. The exponent of $H$ will be left unchanged ($b=0.2$) for convenience of visualization (there are many possible ways to draw these plots), and $R_{init}$ will vary for each diagram, from 0.05 to 0.1. The phase that is obtained through the combination of each $a$-$J$ is indicated by a capital letter inside the phase zone, while the letter symbolizes the phase as was previously described.

For convenience, if a network begins with $Z_k$, $Z_{k+1}$, ..., $Z_4$ being empty, then we would regard the terminal phase as if it has dynamically exhausted these fractions, even if in fact no new revolutionaries had been included, and the network was stationary. For example, if we have $R_{init}=0.001$ and $N_{tot}=90000$ then in the average case we will have $Z^4=Z^3=Z^2=0$, and so phases A, B, C and D overlap. In this case we'll describe the phase as D, to avoid ambiguity.

To begin with, we get very clear phase transitions in all diagrams. All of the phases A-G (except for the E phase) are present through the diagrams.

Phase E is not possible in our networks, as they are fully connected and do not possess unlinked components. Since we begin with $N_s \geq 1$, the conditions allowing all individuals who have at least one revolutionary neighbor to join the revolution would at the very least proliferate the revolution's growth as expanding revolutionary clusters, initiated at each $N_s$ seed, arriving ultimately at phase F (total revolution).

The only visible difference between the diagrams generated for the square lattice and the random graph is in the "splitting of" phase F into phase D for $R_{init}=0.1$. The explanation is that for a square lattice and a sufficiently large fraction of $Z_2$, conditions allowing phase D would result in a total revolution following the dynamics of a bootstrap-percolation [16], [17], [18] rather than our more specific revolution/percolation model. For a random graph, which has a different geographical representation of clusters than a square lattice, this is not the compulsory behavior of the network.

We also note that the phase separation is often portrayed via straight lines, especially for low diversity of $h_i$ (large $c$), and for low $R_{init}$. For more diverse $h_i$ and for low values of $a$ and $J$, the lines are not very straight and the picture is more complex.

Our next goal is to understand these lines, since they are instrumental in the prediction of the revolution's outcome.

From Figure 4 we see that $a^* = R_{init}^{-b}$ presents an important point for all $R_{init}$ and $c$. $a^*$ represents the minimal $a$ at which we may get phase D. When $c \to \infty$ then according to condition (3b) $a^*=a_{min}$, and so we do not witness revolutions for lower $a$ at all.

Let us focus now on the random graph model instead of the square lattice. The results and the theoretical reasoning is the same for both models, while the only difference is in that the square lattice undergoes a very specific dynamics for growing $R_{init}$, which will be later discussed.

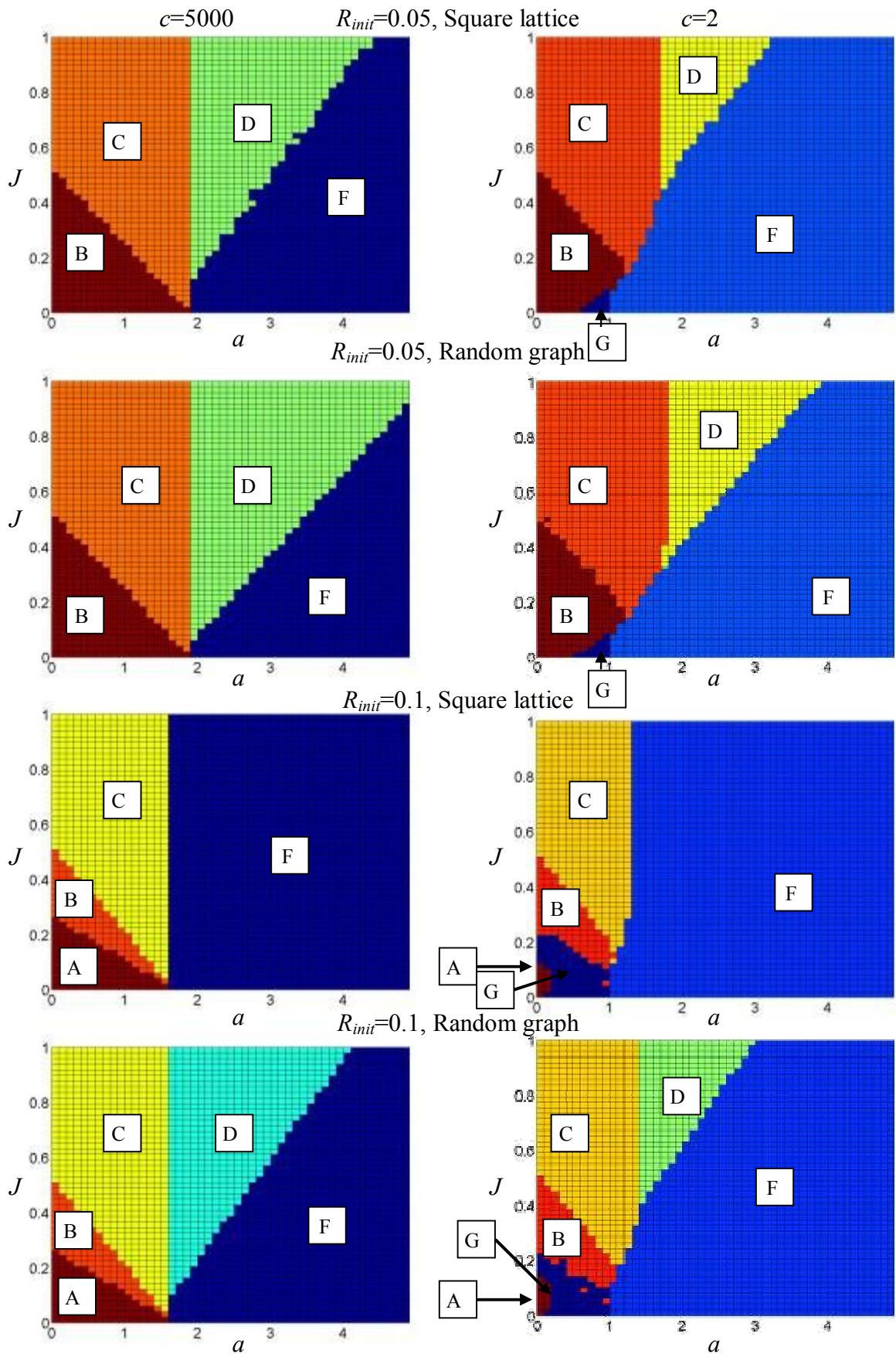

Figure 4: Phase diagrams for a square lattice and a random graph model (with all nodes of degree 4). $N_{tot}$=90000, $b$=0.2 and $c$=5000 or $c$=2.

If we study the graphs for $a>a^*$ we see that the only two possible phases are D (all individuals having at least two revolutionary neighbors join the revolution) and F (total revolution). Therefore, if individuals with less than two revolutionary neighbors join the revolution, we obtain a phase shift and reach total revolution. This is because in this case, the revolution grows first by joining of individuals from $Z_k$ for $k=2,3,4$ and when these dynamics reach a steady state, new members from $Z_1$ or $Z_0$ join the revolution, and 'rekindle' its growth.

Since for $a>a^*$ all those in groups $Z_2$, $Z_3$ and $Z_4$ join the revolution, the equation determining whether an individual will join the revolution having only one revolutionary neighbor is:

$$(5) \quad a(R_{init} + Z_2 + Z_3 + Z_4)^b - 2 \cdot J = \min\{h_i\}$$

while $Z_2$, $Z_3$ and $Z_4$ are given by equation (4). In equation (5), $\min\{h_i\}$ represents the most likely minimal $h_i$ of an individual having one revolutionary neighbor, and not included in the initial random choice of $N_s$. It is straightforward that for $c \to \infty$ $\min\{h_i\}=1$. For growing $Z_1$ and smaller $c$ (small $c$ near 1 causes diverse $h_i$), the value of $h_{min}=\min\{h_i\}$ converges to 0 according to the decreasing probability to obtain larger $h_{min}$: $P(h_i < h_{\min}) \approx 1 - (1 - h_{\min}^c)^{Z_1 \cdot N_{tot}}$.

At any rate, we witness a linear relation between $J$ and $a$ which separates phases D and F.

Now we turn to examine the phase separations when $a<a^*$. At first, let us focus on the non-revolutionary phases.

From considerations similar to those leading to equation (5) we may present the equation governing the separation between phases characterized by the minimum number of revolutionary neighbors an individual must have in order to join the revolution, $k$. Since $a<a^*$ phase D is impossible (and likewise the conditions for phase E which trivially leads to phase F), $k$ may take values of 3 and 4 only, while $k=4$ separates phases A and B (or G and B when $c$ is small) and $k=3$ separates phases B and C.

The line of $a=a^*$ separates phases C and D for large enough values of $J$ (and for low enough values of $R_{init}$ for the square lattice model).

We should state that the results, offered in the following equation are correct for both random graph and the square lattice:

$$(6)\ J = \frac{1}{2\cdot(k-2)} - \frac{\left(R_{init} + \sum_{k'=k}^{4} Z_{k'}\right)^b}{2\cdot(k-2)} a\ ,\ \text{when } k=3\text{ or }4$$

The fact that the initial conditions alone are enough to evaluate which of the three phases (A – or G, B, C) the system will adopt is due to the fact that having individuals from the $Z_4$ and $Z_3$ groups join the revolution makes very little impact on the population, regarding the distributions of $Z_2$, $Z_1$, $Z_0$, and of $R_t$.

It should be added that as seen in Figure 4 for $R_{init}=0.1$ phase A is 'replaced' with phase G as has been previously discussed, and in these cases we are interested in separating G-B rather than A-B

What remains is to analyze the separation of phase F from phases A-G, B and C for $a<a^*$.

Firstly, the minimal value of $a$ at which we may begin to see phase F, $a_{min}$, is obtained for $J=0$. We remind the reader that we have already solved this problem, coming up with

$$(7)\ a_{min} = \max\left(1, R_{init}^{-b+1/c}\right)$$

Moreover, we know that for $a^* = R_{init}^{-b}$ we can obtain $J(a^*)$ by equation (5). If so, the dots in the $J$-$a$ phase diagrams of ($a_{min}$, 0) and ($a^*$, $J(a^*)$) are somehow connected. This connection is not necessarily linear, as can be seen from Figure 4. This implies that a proper causative may not be effective in this case. However, since we have a theoretical evaluation of both $a_{min}$, $a^*$, and $J(a^*)$ we can offer a rather good linear approximation, given by:

$$(8)\ J = \frac{J(a^*)}{a^*-a_{min}}(a - a_{min})$$

A more thorough examination of these predictions and their simulated validation may be found in [15].

This concludes our study of the terminal phase prediction of the revolution for the random graph model with each node of degree 4, and for any combination of network parameters.

However, for the square lattice model and for large enough values of $R_{init}$ phase F may be obtained as a by-product of reaching phase D, as can be seen from the comparison of Figures 4 for $R_{init}=0.1$: we see that the D phase is present only in the random graph model in this case.

Understanding the square lattice is our next step. This is important because the square lattice represents a unique geography of the closest acquaintances, where there is a high probability that the closest acquaintances of any individual may have another common closest acquaintance. This is relevant for some societal structures.

Simulated results show that when $R_{init} \geq R_{init}^{crit}$ for a certain $R_{init}^{crit}$, revolutionary spread becomes independent of $J$ [15], and the governing conditions become again (3a) and (3b), while equations (5) and (7) lose their validity. Moreover, when $R_{init}$ is such that phases D and F overlap, we witness a case of bootstrap-percolation [16] (upon a square lattice). This has been vastly studied, and an approximation for a finite lattice of $LxL$ individuals yields [18] $R^*_{init} \sim \pi^2/[18 ln(L)]$, which allows a percolation (total revolution) with a probability greater than 0.5. For networks having 10,000 to 90,000 individuals as was studied in the current work, $R^*_{init}$ between 0.06 and 0.08 is a robust numerical approximation for the joining of phase D and F on a square lattice.

To complete the analysis we focus on determining the dependence of $R_{init}^{crit}$ on the system parameters $b$ and $c$, and its relation to $R^*_{init}$.

To do so, we will examine the minimal $R_{init}$ which allows a revolution for $a$ obtained by conditions (3a) and (3b), while avoiding looking into very large $R_{init}$, since otherwise the term 'revolution' would be poorly defined. The simulated results for $R_{init}^{crit}$ are shown in Figure 5. These results are product of full simulations, and do not come from just applying conditions (3a) and (3b).

Moreover, some universality is present, in that results depend on the product $bc$ rather than on $b$ and $c$ separately.

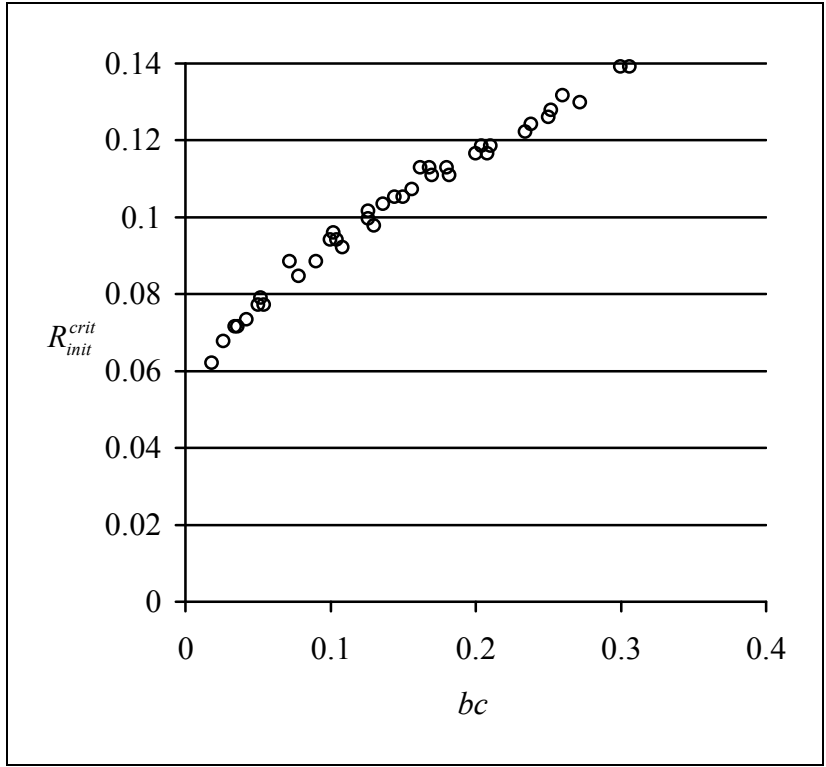

Figure 5: Numerically calculated $R_{init}^{crit}$ for square lattice as a function of $bc$, for networks of different $b$ and $c$: $0.01 \leq b \leq 0.2$, $1 \leq c \leq 6$. Size of $N_{tot}$ was varied as well between 10000 to 90000, but results were the same apart from statistical fluctuations.

We see that as $b$ and/or $c$ increase, $R_{init}^{crit}$ increases as well. This is to be expected, since with growing $b$ the strength of the global field diminishes, and with growing $c$ the inclination of individuals to joining the revolution fades away as well. It is interesting to note that the dependence of $R_{init}^{crit}$ is on the product of the two powers, and not on any other function. Our efforts of theoretically evaluating $R_{init}^{crit}(b \cdot c)$ were unsuccessful, but this calculation is not crucial in determining the outcome of a revolution if we rely on the numerical approximation in Figure 5.

We see that if we limit our research of the square lattice to $R_{init} \leq 0.1$ then only rather low values of $b$ and $c$ will produce $R_{init}^{crit}$ which answers to that condition.

Moreover, we see that for most combinations of $b$ and $c$ we get $R_{init}^{crit} > R_{init}^{*}$, meaning, that for the square-lattice networks which we have studied, phase D effectively causes a total revolution through bootstrap percolation dynamics rather than through the global revolutionary field. This occurs for lower values of $R_{init}$ than those required to have a global revolutionary field producing the same effect.

This is a complete solution of our initial goal of developping a capability to predict the outcome of a revolution, through employing our minimal model.

## 1.5 Conclusions

Besides offering a descriptive view into historical events, and verifying them with our work (which was purely theoretical), we have developed a basic tool for predicting the outcome of a revolutionary venture.

Regarding revolutionary success, as could be derived from our analyses:

- Our model, in contrast to many earlier diffusion models, has a natural range in which revolutions fail (Figures 2 and 4).

- In the failure phases, one may design a "sweet-spots-path" [19] of subsidies schedule that brings the system in the self-sustained, autonomously propagating phase with minimal investment (equation 5).

- The stronger is the effect of the personal relations on individual opinion, the stronger the revolutionary message must be in order to obtain a total revolution, since initially the general population is against revolution (equations 5 and 7, Figure 4).

- From another angle, a single revolutionary may cause a total revolution if the message is strong enough (large $a$, small $b$). This, evidently, is not a new insight, as it is known that all politicians dedicate great funds to their election campaigns, which do exactly that (equations 5 and 7).

- The diversity of opinion (achieved by small $c$ near unity) towards the revolutionary endeavor is a significant factor in determining its success. Therefore, the revolutionary message should focus not only on attracting the least public resistance, but also on stirring multiplicity of views on the subject (condition (3b), Figure 4).

- Having a society where the most influential people in any person's life are intertwined, as is simplified by the square lattice model in contrast to the random graph, may enhance the revolutionary spread, but only when the initial fraction of revolutionaries is rather large (Figure 5). For small revolutionary seeds, the geography of the social network is immaterial to the effects of the revolution (Figure 4).

- As one varies the strength of the revolutionary message and the initial number of exogenously induced supporters, one encounters discontinuous systemic transitions from failure to total revolution (Figure 4).

- The clear-cut phase transitions we have revealed imply that sometimes, a small advance in enhancing the revolutionary message power ($a$) may drastically affect the outcome, i.e. cause a phase shift (Figure 4). Moreover, the larger the initial revolutionary seed, the more significant this phase shift is: leaping from phase C to F, or even from phase C to D translates into a large (or even total) addition of people to the revolution (equations 5, 6, 7).

### Appendix: Spontaneous Revolutions

It is proper to examine the possibility of a spontaneous revolution, without having an initial constellation of revolutionary catalysts throughout the population.

This procedure could be modeled via heat-bath dynamics at temperature $T$ below the zero-field critical temperature $T_c$.

In this appendix, we compose a very basic analysis, for square lattice networks of varying size.

Explicitly, the chance of any individual $i$ to join the revolution (attain the value of $S_i$=-1) is:

$$(a1) \quad P(S_i = -1) = \frac{e^{-2E/T}}{1+e^{-2E/T}},$$

where, $E = h_i - H + H_{local}(i)$, or $E = h_i - aR^b + J\sum_{j=1}^{4} S_j$, from [14] and many other works.

Thus, even if the network is initially revolutionary-free, the finite temperature imposes a probability that individuals will randomly join the revolution according to equation (a1). This random joining of individuals will enhance the global field $H$ and consequently increase the probability of further individuals join.

Before introducing finite temperature, the joining of an individual with the revolution was manifested through a simple inequality: $h_i < H - H_{local}(i)$, which once reached, would remain true until steady state. With finite temperature however, this is not necessarily the case, and we may allow individuals to leave the revolution with the probability of $P(S_i=1)=1-P(S_i=-1)$, according with equation (a1).

If we assume that an individual joins a revolution 'forever', then the revolution is achieved by an Arrhenius law as a function of temperature, as presented in Figure a1, for varying lattice sizes, $LxL$:

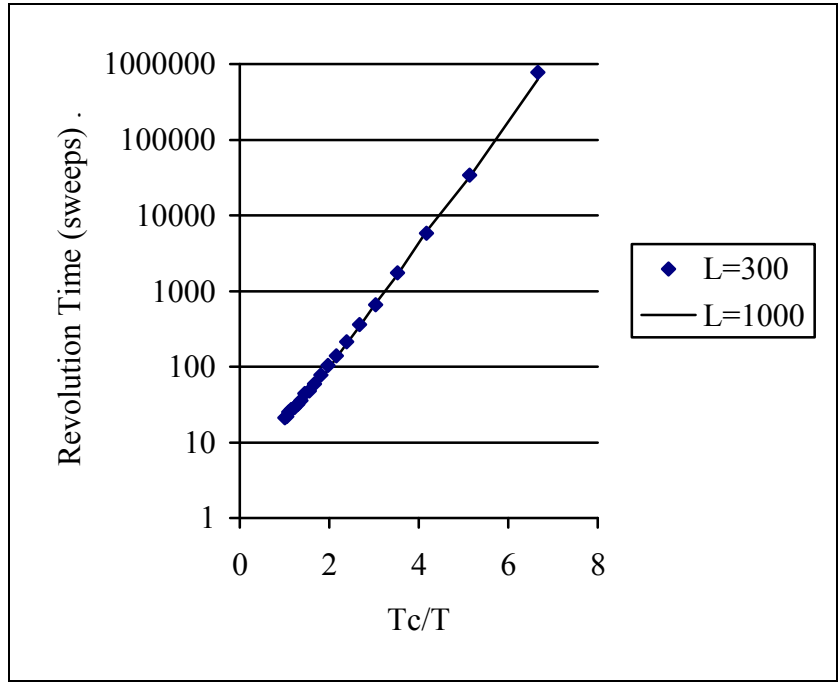

Figure a1: The time at which a total revolution was reached, measured in the number of lattice sweeps (= times each individual was visited). We see an Arrhenius law. Parameters used: $J=1$, $a=1$, $b=1$, $c=1$.

If we do allow individuals the chance of leaving the revolution, then we should redefine our meaning of 'revolution', since our previous requirement had all of the individuals join in, and would now be unrealistically improbable. If so, we may claim that a revolution is obtained when the number of revolutionaries exceeds the number of old-rule supporters, while the network is initialized without a single revolutionary. Figure a2 shows this newly defined revolution time as a function of temperature:

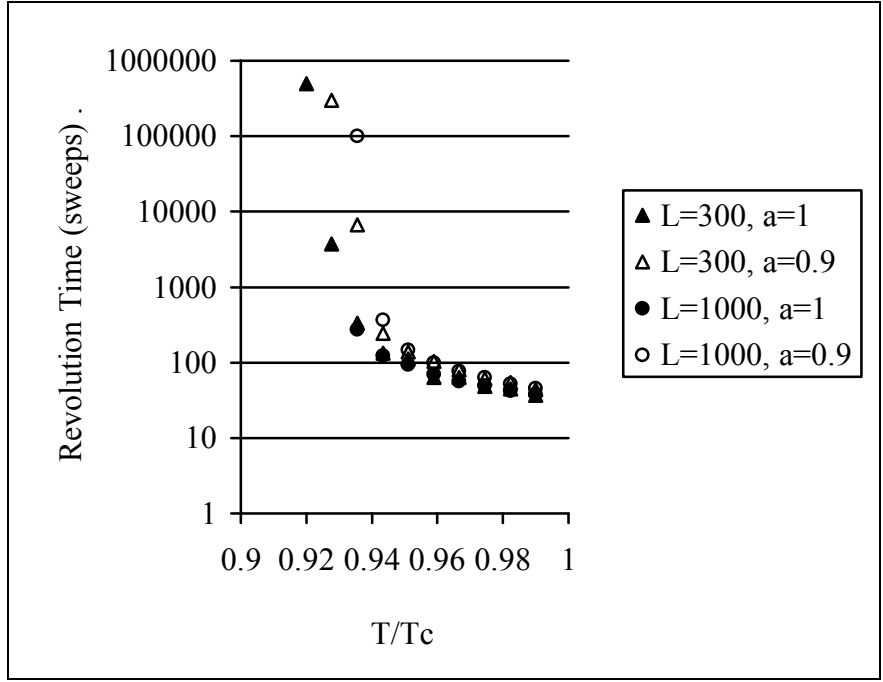

Figure a2: Parameters used: $J=b=c=1$, $a=1$ and $a=0.9$, for lattices of $L=300$ and $L=1000$.

Delving further into analyzing the dependence of the revolution time on the temperature is beyond the scope of this work. We can preliminarily state that the above results deviate from the Vogel-Fulcher for the relaxation (i.e. revolution) time of $\tau \sim e^{\frac{A}{(T/T_c - B)}}$.